\documentclass{jetpl}
\twocolumn
\lat
\usepackage{cite}
\usepackage{graphicx}
\usepackage{url}

\title{The complex singularity of a Stokes wave}
\rtitle{The complex singularity of a Stokes wave}
\sodtitle{The complex singularity of a Stokes wave}
\author{S.\,A.~Dyachenko$^{+}$,  P.\,M.~Lushnikov$^{+*}$, and A.\,O.~Korotkevich$^{+*}$\/\thanks{e-mail: alexkor@math.unm.edu}}
\rauthor{S.\,A.~Dyachenko, P.\,M.~Lushnikov, and A.\,O.~Korotkevich}

\sodauthor{Dyachenko, Lushnikov, Korotkevich}

\address{$^+$Department of Mathematics and Statistics, University of New Mexico, MSC01 1115, 1 University of New Mexico, Albuquerque, NM 87131-0001, USA\\~\\
$^*$L.\,D.~Landau Institute for Theoretical Physics, 2 Kosygin Str., Moscow, 119334, Russian Federation}

\dates{6 November 2013}{*}
\abstract{Two-dimensional potential flow of the ideal  incompressible fluid with  free surface and infinite depth  can be described by a conformal map of the fluid domain into the complex lower half-plane.
 Stokes wave  is  the fully  nonlinear gravity wave propagating with the constant velocity. 
The increase of the scaled wave height $H/\lambda$ from the linear limit $H/\lambda=0$ to the critical value $H_{max}/\lambda$ marks the transition from the limit of almost linear wave to a strongly
nonlinear limiting Stokes wave. Here $H$ is the wave height  and $\lambda$ is the wavelength. We simulated  fully nonlinear Euler equations, reformulated in terms of conformal variables, to find Stokes
waves for different wave heights.
Analyzing spectra of these solutions we found in conformal variables, at each Stokes wave height, the distance $v_c$ from the lowest singularity in the upper half-plane  to the real line which corresponds to the fluid free surface.
We also identified that this singularity is the square-root branch point.
The limiting Stokes wave emerges as the singularity reaches the fluid surface. From the analysis of data for  $v_c\to 0$ we suggest a new power law scaling $v_c\propto (H_{max}-H)^{3/2}$
as well as new estimate $H_{max}/\lambda \simeq 0.1410633$.
}
\PACS{02.60Cb, 47.10.-g, 47.35.Bb, 92.10.Hm}
\begin{document}

\maketitle

Theory of spatially periodic progressive (propagating with constant velocity without change of the amplitude) waves
in two-dimensional (2D) potential flow of an ideal incompressible fluid
with free surface in gravitational field was founded in pioneering works by Stokes~\cite{Stokes1847,Stokes1880} and developed further by
Michell~\cite{Michell1893}, Nekrasov~\cite{Nekrasov1921,Nekrasov1951}, and many others (see e.g. a book
by Sretenskii~\cite{Sretenskii1976} for review of older works as well as Refs.
\cite{MalcolmGrantJFM1973LimitingStokes,SchwartzJFM1974,Longuet-HigginsFoxJFM1977,Williams1981,WilliamsBook1985,CowleyBakerTanveerJFM1999,Longuet-HigginsWaveMotion2008,BakerXieJFluidMech2011.pdf} and Refs. there in for more recent progress). There are two major approaches to analyze the Stokes wave, both originally  developed by Stokes.
The first approach is the
perturbation expansion in amplitude of Stokes wave called by the Stokes expansion. That approach is very effective for small amplitudes but converges very slowly (or does not converge at all, depending on the formulation) as the wave approaches to the maximum height $H_{max}$ (also called by the wave of the greatest height or the limiting Stokes wave) which is defined at the distance from the crest to the trough of Stokes wave over a spatial period $\lambda$.
The second approach is to consider a limiting Stokes wave, which is the progressive wave with the highest nonlinearity. Stokes found that the limiting
Stokes wave has the sharp angle of $2\pi/3$ radians on the
crest~\cite{Stokes1880sup}, i.e. the surface is non-smooth (has a jump of slope) at that spatial point. That corner singularity explains a slow convergence of Stokes expansion
as $H\to H_{max}.$

\begin{figure*}[ht!]
\centering
\includegraphics[width=6.8in]{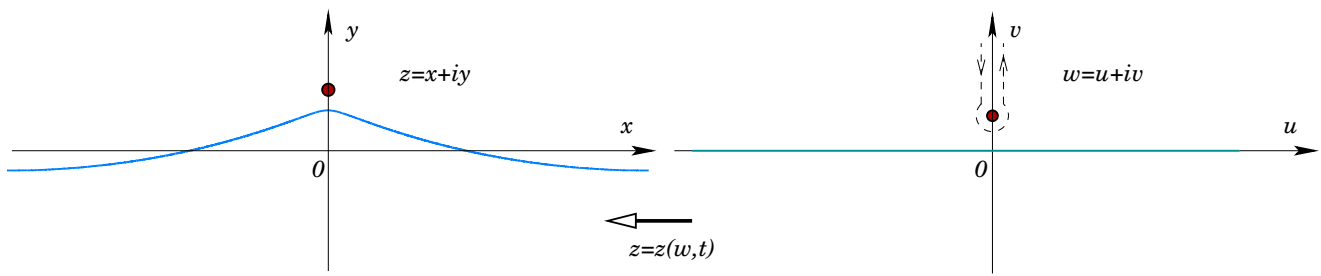}
\caption{Fig.~\ref{conformal_map}.
Schematic of a conformal map between  the domain  below the solid line (left panel) in $z=x+iy$ plane and the lower half-plane in $w=u+iv$ (right panel).  Fluid occupies the domain below the solid line in physical plane  $z=x+iy.$  The solid line of  left panel (corresponds to a free surface of the fluid) is mapped into the real line (another solid line) in  right panel. One spatial period of Stokes wave is shown by solid lines in both panels in the reference frame moving with the velocity $c$.  The dark circles mark the positions of the  singularity closest to the fluid surface in both panels. The dashed line in right panel  shows the integration contour as it is moved  from the real line upward, see text for more details.  }
\label{conformal_map}
\end{figure*}

It was Stokes~\cite{Stokes1880sup} who proposed to use conformal mapping in order to address
 finite amplitude progressive wave.
In this letter we consider the particular case of infinite depth fluid although  more general case of fluid
of arbitrary depth can be studied in a similar way. Assume that the free surface is located at $y=\eta(x,t)$, where $x$ is the horizontal coordinate, $y$ is the vertical coordinate, $t$ is the time
and $\eta(x,t)$ is the surface elevation with respect to the mean level of fluid, i.e. $\int^\infty_{-\infty}\eta(x,t)dx=0$.
We consider the conformal map of the domain $-\infty<y< \eta(x,t), \ -\infty<x<\infty$  of the complex plane $z\equiv x+iy$ filled by the infinite depth fluid into a lower complex half-plane
(from now on denoted by $\mathbb{C^-}$) of a variable $w\equiv u+iv$ (see Fig.~\ref{conformal_map}). The free surface is mapped into the real line $v=0$ with $z(w)$ being the analytic function in the lower half-plane of $w$ as well as the complex fluid potential $\Pi(w)$ is also analytic in  $\mathbb{C^-}$.
Both $z(w)$ and $\Pi(w)$ have singularities in upper half-plane (here and further denoted by $\mathbb{C}^+$). The  knowledge of singularities in $\mathbb{C}^+$ would result in the efficient description of the solution in the physical variables. Examples of such type of solutions in hydrodynamic-type systems are the dynamics of free
surface of ideal fluid with infinite depth
\cite{KuznetsovSpektorZakharovPRE1994}  and finite depth
\cite{DyachenkoZakharovKuznetsovPlasmPhysRep1996}, dynamics of interface
between two ideal fluids \cite{KuznetsovSpektorZakharovPhysLett1993},
ideal fluid pushed through viscous fluid in a narrow gap between
two parallel plates (Hele-Shaw flow) \cite{MineevZabrodinPRL2000} and the dynamics of the interface between ideal  fluid and light viscous
fluid   \cite{LushnikovPhysLettA2004}. In all these systems the dynamics is determined by  poles/branch cuts in the complex plane.

In this Letter we determine that for Stokes wave the lowest singularities in  $\mathbb{C}^+$ of both $z(w)$ and $\Pi(w)$  are the square-root branch points located   periodically at  $w=n\lambda+i v_c-ct, \ n = 0, \pm 1, \pm 2,...$  (we choose that the crests of Stokes wave to be located at $w=n\lambda-ct$) and we determine $v_c$ numerically as a function of $H/\lambda$. Here $c$ is the velocity of propagation of Stokes wave  which depends on $H$. We found that as $H\to H_{max}$, the branch point approaches real axis with the scaling law
\begin{align} \label{vcscalinglaw}
v_c \propto (H_{max}-H)^{\delta},
\end{align}
where $\delta=1.48\pm 0.03$. At $H=H_{max}$ the branch point reaches the real axis producing the corner singularity at the free surface of limiting Stokes wave. We believe that this new scaling law provides the efficient way for the description of near-limiting Stokes waves. Adiabatically slow approach of Stokes wave to its limiting form during wave dynamics is one of the possible routes to wave breaking and whitecapping, which are responsible for significant part of energy dissipation for gravity waves~\cite{ZKPR2007, ZKP2009}. Formation of a close to limiting Stokes wave is also considered to be a probable final stage of evolution of a freak (or rogue) waves in the ocean resulting in formation of approximate limiting Stokes wave for a limited period of time with following wave breaking and disintegration of the wave or whitecapping and attenuation of the freak wave into wave of regular amplitude \cite{ZakharovDyachenkoProkofievEuropJMechB2006,RaineyLonguet-HigginsOceanEng2006}.

{\it Basic equations.} In physical coordinates $(x,y)$ a velocity ${\bf v}$   of $2D$ potential flow of inviscid incompressible fluid is determined by a velocity potential $\Phi$ as ${\bf v}= \nabla \Phi$.
The incompressibility condition $\nabla \cdot {\bf v} = 0$  results in the Laplace equation
\begin{align} \label{laplace}
\nabla^2 \Phi = 0
\end{align}
inside fluid $-\infty<y<\eta(x,t)$. To obtain the closed set of equations we add the decaying boundary condition at large depth $\Phi(x,y,t)|_{y\to-\infty} = 0$, the kinematic boundary condition
\begin{align} \label{kinematic1}
\dfrac{\partial \eta}{\partial t}    =\left( -\dfrac{\partial \eta}{\partial x}\dfrac{\partial \Phi}{\partial x}+
\left.\dfrac{\partial \Phi}{\partial y}\right)\right|_{y = \eta(x,t)}
\end{align}
and the dynamic boundary condition
\begin{align} \label{dynamic1}
\left.\left(\dfrac{\partial \Phi}{\partial t} +
 \dfrac{1}{2}\left(\nabla \Phi\right)^2\right)\right|_{y = \eta(x,t)} + g\eta = 0
\end{align}
at the free surface $y=\eta(x,t).$ We  define the boundary value of the velocity potential as $\left.\Phi(x,y,t)\right|_{y = \eta(x,t)} \equiv \psi(x,t)$.

The system \eqref{laplace}-\eqref{dynamic1}   was recast into the conformal variables   first in Ref.~\cite{Ovsyannikov1973} and later
independently in Ref.~\cite{DKSZ1996} taking the following form \cite{DKSZ1996}:
\begin{equation}\label{fullconformal1}
 y_tx_u  -x_t  y_u + \hat H \psi_u = 0
\end{equation}
for the kinematic boundary condition and
\begin{equation}\label{fullconformal2}
\psi_t y_u - \psi_u y_t + gyy_u =- \hat H \left (\psi_t x_u - \psi_ux_t + gyx_u \right )
\end{equation}
for the dynamic boundary condition. Here $\hat H f(u)=\frac{1}{\pi} P.V.
\int^{+\infty}_{-\infty}\frac{f(u')}{u'-u}du' $ is the Hilbert
transform with $P.V.$ meaning a Cauchy principal value of integral.
Equations  \eqref{fullconformal1} and \eqref{fullconformal2} are equivalent to  \eqref{laplace}-\eqref{dynamic1} because $z(w)$ is assumed to be analytic in $\mathbb{C}^-$. Both \eqref{fullconformal1} and \eqref{fullconformal2} are defined on the real line $w=u$.
Note that a Fourier transform of $\hat H$ results in the multiplication operator $ (\hat H f)_k=i\, \text{sign}{\,(k)}\,f_k. $ The complex potential $\Pi$ is recovered from the analytical continuation of $(1+i\hat H)\psi$ into $\mathbb{C}^-. $ Also, the analyticity of $z$ in $\mathbb{C}^-$ implies that
\begin{align} \label{xytransform}
y=\hat H\tilde x \quad  \text{and} \quad  \tilde x=-\hat Hy,
\end{align}
where $\tilde x(u,t)\equiv x(u,t)-u$ and  $\tilde z(u,t)\equiv z(u,t)-u.$
We fix the location of fluid surface in $y$ by the condition that the mean elevation $\langle y(1+\tilde x_u)\rangle=0$, where $\langle \ldots \rangle$ stands for the average in $u$.

{\it Progressive waves.}
Stokes wave corresponds to a solution of system \eqref{fullconformal1} and \eqref{fullconformal2} in the travelling wave form
\begin{align}\label{travelling}
\psi (u,t)= \psi(u-ct), \
\tilde z (u,t)= \tilde z(u-ct),
\end{align}
where both $\psi$ and $\tilde z$ are the periodic functions of $u-ct$.  We transform into the moving frame of reference, $u-ct\to u$,
and assume that the crest of the Stokes wave is located at $u=0$ as in Fig. \ref{conformal_map} and $L$ is the period in
$u$ variable for  both $\psi$ and $\tilde z$ in \eqref{travelling}.
The Stokes solution requires  $y(u)$ to be the even function while $\tilde x(u)$ needs to be the odd function.
Taking into account the periodicity of $\tilde x(u)$ in \eqref{travelling} it implies that $\tilde x(\pm L/2)=0$. Then
 the spatial period of the Stokes solution  is the same, $L=\lambda$, both in $x$ variable (i.e. for $\eta(x-ct)$) and  $u$ variable (i.e. for \eqref{travelling}).

It follows from \eqref{fullconformal1} and \eqref{travelling} that    $\psi=-c\hat H y$ and then excluding $\psi$ from \eqref{fullconformal2} we obtain that

\begin{equation}
\label{stokes_wave}
-c^2y_u + gyy_u + g\hat H[y(1+\tilde x_u)] = 0
\end{equation}
We now apply $\hat H$ to \eqref{stokes_wave}, use \eqref{xytransform} to obtain a closed expression for $y$, and introduce the operator   $\hat k \equiv -\partial_u \hat H = \sqrt{-\nabla^2}$ which results in the following expression
\begin{equation}\label{stokes_wave2}
\begin{split}
& \hat L_0 y\equiv \left(  \dfrac{c^2}{c_0^2}\hat k - 1 \right) y -  \left( \frac{\hat k y^2}{2} + y\hat k y \right) = 0, \\
\end{split}
\end{equation}
where $c_0 = \sqrt{g/k}$ is the phase speed of linear gravity wave with $k=2\pi/\lambda$ and we made all quantities dimensionless by the following scaling transform $u\to u\lambda/2\pi,\ x\to x\lambda/2\pi, \ y\to y\lambda/2\pi $. In these scaled units the period of  $\psi$ and $\tilde z$  is $2\pi.$

We solve  \eqref{stokes_wave2} numerically to find $y(u)$ by two different methods each of them beneficial for different range of parameters. First method is inspired by a Petviashvili
method~\cite{Petviashvili1976} which was originally proposed to find solitons in nonlinear Schrodinger (NLS) equation as well as it was adapted for nonlocal NLS-type equations, see e.g. \cite{LushnikovOL2001}. Here we  use
a generalized Petviashvili method  (GPM) ~\cite{LY2007}  designed to solve the general equation
$\hat L_0 y \equiv-\hat My+\hat Ny= 0$
for the unknown function $y(u)$, where $\hat L_0$ is the general  operator which  includes a linear part $-\hat M$ and a nonlinear part $\hat N$. For our particular form of $\hat L_0$ in  \eqref{stokes_wave2} we have that $-\hat M=  \left(  \dfrac{c^2}{c_0^2}\hat k - 1 \right) $.
$\hat M$ is invertible on the space of $2\pi$-periodic functions because  Stokes wave requires $1<c^2/c_0^2< 1.1$ \cite{WilliamsBook1985}. The convergence of GPM~\cite{LY2007,PelinovskyStepanyantsSIAMNumerAnal2004} is determined by the smallest negative eigenvalue of the operator $\hat M^{-1}\hat L$. Here $\hat L$ is the linearization operator of  $\hat L_0$ about the  solution $y$ of \eqref{stokes_wave2}: $\hat L \delta y=-\hat M\delta y-  \left( \hat k (y\delta y) + y\hat k \delta y+\delta y\hat k y \right) $.
It is assumed that $\hat M^{-1}\hat L$ has only a single positive eigenvalue $1$ determined by $
\hat Ly = \hat My
$. GPM iterations are given by  \cite{LY2007} \begin{equation}
\label{GP}
y_{n+1} - y_{n} = \left( \hat M^{-1}L_0y_n -
\gamma\dfrac{\langle y_n,  \hat L_0y_n \rangle}{\langle y_n, \hat My_n \rangle}y_n  \right)\Delta \tau,
\end{equation}
where  $\Delta \tau>0$ is the  parameter
that controls a  convergence speed of iterations and  $\gamma = 1 + \frac{1}{\Delta \tau}$ is chosen to project iterations into the subspace orthogonal to $y$
(the only eigenfunction $y$ with the positive eigenvalue). In practice this method allowed  to find high precision solutions  up to
$H/\lambda \lesssim  0.1388 $ as GPM
requires significant decrease of  $\Delta \tau$ with the growth of $H/\lambda$ to have convergence.
For larger $H/\lambda$ we used the second method which is the Newton
Conjugate Gradient  (Newton-CG) ~method\cite{JiankeYang2009,YangBook2010}.
The idea behind the Newton-CG method is simple and aesthetic: firstly, linearize  \eqref{stokes_wave2} about the current approximation $y_n$:
$ \hat L_0y_n + \hat L_1\delta y   = 0, $ where $\hat L_1=-\hat M\delta y-  \left( \hat k (y_n\delta y) + y_n\hat k \delta y+\delta y\hat k y_n \right)$ - linearization of $\hat L_0$ on the current approximation $y_n$.
Secondly, solve the resulting linear system for $\delta y$ with one of your
favourite numerical methods, in our case it was either Conjugate Gradient (CG) method or Conjugate Residual (CR) method~\cite{Luenberger1970} to obtain next approximation $y_{n+1}=y_n+\delta y$. It should be noted that monotonic
convergence of CG or CR methods are proven only for positive definite (for CR -- semidefinite) operators, while in our
case $\hat L_1$ is indefinite. Nevertheless, both methods were converging to the solutions, and convergence was much
faster than using GPM.

Newton-CG/CR methods can be written in either Fourier space, or in physical space. We considered both cases, however
Newton-CG/CR methods in Fourier space require four fast Fourier transforms per CG/CR step, while in physical space it
requires at least six. 
For both cases CG and CR we used preconditioner $\hat M$.

We found that the region of convergence of the  Newton-CG/CR methods  to nontrivial physical solution  \eqref{stokes_wave2} is quite narrow and requires  an   initial guess $y_0$ to be very close to the exact solution $y$. In practice we first run GPM and then  choose $y_0$ for Newton-CG/CR methods as the last available iterate
of GPM.
%
%
\begin{figure}[ht!]
\centering
\includegraphics[width=3.2in]{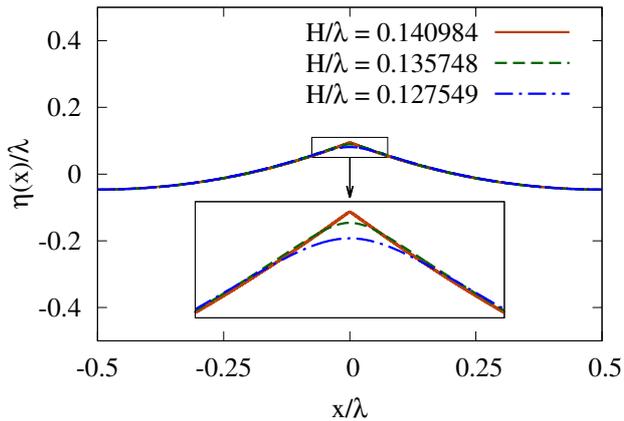}
\caption{\label{many_waves}Fig.~\ref{many_waves}. Stokes wave for $c/c_0=1.082500$ (blue line), $c/c_0=1.091500$ (green line) and $c/c_0=1.0922795$ (dark orange line). Corresponding values of $H/\lambda$ are given in the legend. Inset shows zoom-in into small values of $x/\lambda$ near a wave crest. }
\end{figure}

{\it Finding singularity  from Fourier spectrum.} Assume that the singularity of $\tilde z$ closest to real axis in $w$ complex plane is  the branch point %
\begin{align} \label{ztildebranchpoint}
\tilde z\simeq c_1(w-i v_c)^\beta
\end{align}
 for $w\to i v_c$, where $c_1$ is the complex constant, $v_c>0$ and $\beta$ are real constants.  By the periodicity in $u$, similar branch points are located at $w=i v_c+2\pi n, \, n=\pm1,\pm 2,\ldots$ We expand $\tilde z(u)$ in Fourier series $\tilde z(u)=\sum\limits_{k=0}^{-\infty} \hat {\tilde  z}_k\exp[iku]$, where
\begin{align} \label{ztildefourier}
\hat {\tilde  z}_k=\frac{1}{2\pi}\int\limits_{-\pi}^\pi \tilde z(u) e^{-iku}du
\end{align}
is the Fourier coefficient and the sum is taken over nonpositive integer values of $k$ which ensures both $2\pi$-periodicity of $\tilde z(u)$ and analyticity of $\tilde z(w)$ in $\mathbb{C}^-$. We evaluate \eqref{ztildefourier} in the limit $k\to -\infty$ by moving the integration contour from $-\pi<u<\pi$ into $\mathbb{C}^+$  until it hits the lowest branch point \eqref{ztildebranchpoint} so it goes around branch point and continues straight upwards about both side of the corresponding branch cut as shown by the dashed line in right panel of Fig. \ref{conformal_map}. Here  we assume that branch cut is a straight line connecting $w=i v_c$ and $+i\infty$. Then the asymptotic of $|\hat {\tilde  z}_k|$ is given by
\begin{align} \label{zkasymptotic}
|\hat {\tilde  z}_k|\propto |k|^{-1-\beta}e^{-|k|v_c}, \quad k\to -\infty.
\end{align}

In simulations we expand $y(u)$ in cosine Fouries series using FFT to speed up simulations. After that we can immediately recover $\tilde  z(u)$ by  \eqref{xytransform}.

We calculated $\tilde  z(u)$ with high accuracy for different values of  $H/\lambda$ using computations in  quad precision (32 digits) to have wide enough dynamic range for Fourier spectrum to recover $v_c$ in \eqref{zkasymptotic} with high precision. Fig. \ref{many_waves} shows the spatial profiles of Stokes waves for several values of $H/\lambda$ in physical variables $(x,y).$ The Stokes wave quickly approaches the profile of limiting wave except a small neighborhood of the crest.

\begin{figure}[ht!]
\centering
\includegraphics[width=3.2in]{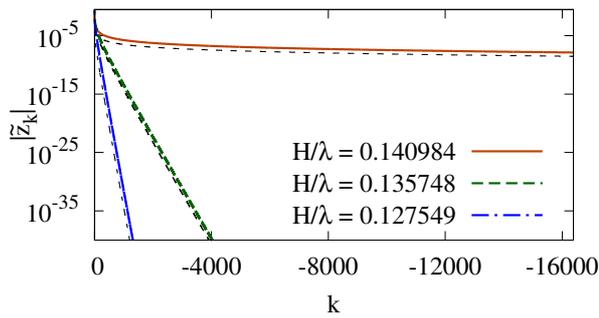}
\caption{\label{wave_spectrum}Fig.~\ref{wave_spectrum}. Spectra of Stokes waves for the values  of $H/\lambda$ as in Fig. \ref{many_waves} (thick lines) and fit to  \eqref{zkasymptotic} with $\beta=1/2$ (thin dashed lines). Solid line shows only a small fraction of the actually numerically resolved spectrum.}
\end{figure}

Fig.~\ref{wave_spectrum} gives examples of Fourier spectra $|\hat {\tilde  z}_k|$ for several values of $H/\lambda$ (solid lines) which are in excellent agreement with the fitted asymptotic \eqref{zkasymptotic}
 (dashed lines) for large $|k|$ provided that $\beta=1/2$, while
other values of $\beta$ give much worse fit. This is consistent with the prediction $\beta=1/2$ in Refs. \cite{MalcolmGrantJFM1973LimitingStokes} and \cite{CowleyBakerTanveerJFM1999}.

\begin{figure}[ht!]
\centering
\includegraphics[width=3.2in]{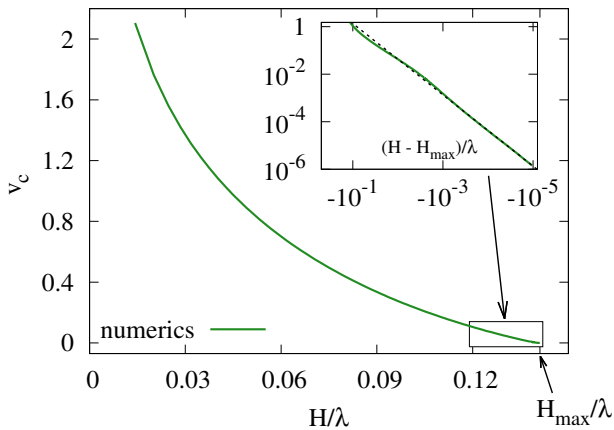}
\caption{\label{Distance}Fig.~\ref{Distance}. Position $v_c$  of  closest singularity of $z(w)$ as a function of $H/\lambda$ obtained numerically. Inset shows zoom-in for $(H_{max}-H)/\lambda\ll 1$ with $H_{max}$ taken from \eqref{Hmaxnew} and dashed line is the power law  \eqref{vcscalinglaw}.  }
\end{figure}

Fig. \ref{Distance} shows the dependence of $v_c$ (in rescaled units given above) on $H/\lambda$ obtained from the numerical fit of spectra to \eqref{zkasymptotic} with $\beta=1/2$. It is seen that the approach of the branch point $w=iv_c$ to the real axis slows down with the increase of $H/\lambda$.
The number of Fourier modes which we used in FFT for each value $H/\lambda$ increases quickly with the increase of $H$ as $v_c$ decreases. E.g., for $H/\lambda=0.0994457$ it was enough to use 256 modes while for the largest wave height
\begin{align} \label{Hmaxnum}
H_{max}^{num}/\lambda=0.1410532464103
\end{align}
achieved in simulations we used $2^{25}\approx 32\times10^{6}$ modes. That extreme case has $c/c_0=1.092285125000$ and  $v_c=1.42800  \cdot 10^{-6}\pm10^{-11}$.

The best previously available estimate of $H_{max}$ was found in Ref. \cite{WilliamsBook1985} as $H^{Williams}_{max}/\lambda= 0.141063.$ However, it was not fully accepted by the community lacking an independent confirmation. Instead the other  commonly used but less precise estimate is  $H^{Schwarts}_{max}/\lambda= 0.1412$ \cite{SchwartzJFM1974}. Fig.~\ref{Distance} shows our numerical values of $v_c$ in the limit  $(H_{max}-H)/\lambda\ll 1$ fitted to   the scaling law  \eqref{vcscalinglaw} with
\begin{align} \label{Hmaxnew}
H_{max}/\lambda=0.1410633\pm 4\cdot10^{-7}.
\end{align}
The mean-square error for $\delta$ in \eqref{vcscalinglaw} is  $\simeq 0.04$ which offers the exact value $\delta=3/2$ as a probable candidate for  \eqref{vcscalinglaw}.
Our new estimate \eqref{Hmaxnew} suggests that the previous estimate $H^{Williams}_{max}$ is more accurate than $H^{Schwarts}_{max}$.  The difference between   \eqref{Hmaxnew}  and the  lower boundary \eqref{Hmaxnum} of the largest $H$  is $\simeq0.007\%$.
However, we would like to point out that two independent fits required to obtain \eqref{Hmaxnew} made our estimate accuracy in  \eqref{Hmaxnew} not very reliable while the lower bound  \eqref{Hmaxnum} is very reliable and obtained with very high precision.


%
%

In summary, we found the Stokes solutions of the primordial Euler equations with free surface for large  range of wave heights, including the approach to
the limiting Stokes wave. Through the analysis  of the solution spectra we calculated in the conformal variables the distance of the lowest singular point in the upper complex half
plane to the surface of the fluid as a the function of $H$. We found that   this singularity is the square-root branch point.
The limiting Stokes wave emerges as the singularity reaches the fluid surface. We found from our high precision simulations the lower bound \eqref{Hmaxnum} for the limiting wave of the greatest height $H_{max}$. We also fitted $v_c(H)$ to the scaling law
\eqref{vcscalinglaw} which suggests that it might be exactly $\delta=3/2$ as well as it provides the new estimate  \eqref{Hmaxnew} for $H_{max}.$

The authors would like to thank A.\,I.~Dyachenko for fruitful discussions on application of Petviashvili method to the dynamical
equations in conformal variables and D.\,Appel\"o for discussion about CG and CR methods.
Work of S.D. and A.K. were partially supported by the NSF grant OCE 1131791. Also the authors would like to thank developers of FFTW~\cite{FFTW} and the whole GNU project~\cite{GNU} for developing, and supporting this useful and free software.

\bibliographystyle{h-physrev}

\end{document}